\documentclass[review]{elsarticle}

\usepackage{lineno,hyperref}
\modulolinenumbers[5]

\journal{Journal of Informetrics}

\usepackage{amsmath,amssymb,amsfonts}
\usepackage{url}
\usepackage{multirow}
\usepackage{threeparttable}
\usepackage{booktabs}
\usepackage{graphicx}
\usepackage{subfigure}
\usepackage{comment}
\usepackage{caption}
\usepackage{enumerate}
\usepackage{fancyhdr}
\begin{document}

\begin{frontmatter}

\title{The statistical nature of $h$-index of a network node and its extensions}

\author[dlu]{Yan Liu}
\author[mymainaddress]{Mudi Jiang}
\author[mymainaddress]{Lianyu Hu}

\author[mymainaddress,mysecondaryaddress]{Zengyou He\corref{mycorrespondingauthor}}
\cortext[mycorrespondingauthor]{Corresponding author}
\ead{zyhe@dlut.edu.cn}

\address[dlu]{School of Software Engineering, Dalian University, Dalian, 116622, China}
\address[mymainaddress]{School of Software, Dalian University of Technology, Dalian, 116024, China}
\address[mysecondaryaddress]{Key Laboratory for Ubiquitous Network and Service Software of Liaoning Province, Dalian, 116024, China}

\begin{abstract}
Evaluating the importance of a network node is a crucial task in network science and graph data mining.
$H$-index is a popular centrality measure for this task, however, there is still a lack of its interpretation from a rigorous statistical aspect.
Here we rewrite $h$-index in a new way based on order statistics, which allows us to obtain a new family of centrality indices by generalizing the $h$-index along this direction.
The theoretical and empirical evidence shows that such a statistical interpretation enables us to obtain a general and versatile framework for quantifying the importance of a network node.
Under this framework, many new centrality indices can be derived and some of which can be more accurate and robust than $h$-index.
We believe that this research opens up new avenues for developing more effective indices for node importance quantification from a viewpoint that still remains unexplored.
\end{abstract}


\end{frontmatter}


\section{Introduction}

Complex networks have been used to portray a variety of interactive systems \cite{albert2002statistical, newman2003structure}, such as traffic networks \cite{LORDAN2014112traffic, LORDAN201618traffic}, social networks \cite{morone2015social, eubank2004social}, power grids \cite{Yangeaan3184power, 5751206power}, neural networks \cite{gurney1997neural, Kennedy1488Neural}, economic networks \cite{JACKSON2002265ecnomic, 1990economic} and many others.
Finding important and influential nodes is of great significance for understanding the network structures and functions.
Therefore, how to effectively and efficiently recognize those vital nodes in a network has become a fundamental research issue in network science \cite{cp2021tudisco,cp2021fan,cp2022evans}.

Many measures have been proposed for quantifying the importance of nodes in a network.
In \cite{lobby2009korn},  the Hirsch's $h$-index \cite{hindex2005hirsch} for evaluating the scientific research output of a scholar is employed to quantify the importance of a network node.
In the network setting, the $h$-index of a node is the maximal value $h$ such that it has at least $h$ first-order neighbors whose degrees are no less than $h$. Since its introduction into the network science societies, the $h$-index has been extensively investigated from different aspects \cite{lu2016h} and extended to weighted networks \cite{weighted_h2019wu}, directed networks \cite{directed_h2018zhai, directed_h2019zhai} and multilayer networks \cite{multilayer2017basaras}.

As one of the de facto standard measures \cite{lu2016vitalreview1}, the $h$-index has been successfully applied to different domains \cite{das2018review2, review32019Liu, review42019Li, review52019bian}.
Furthermore, some methods for extracting the core structure are inspired by $h$-index \cite{core_structure2021wang,core_structure2018zhang,core_structure2014zhao,core_structure2009schubert}.
Unfortunately, how to understand and interpret the $h$-index of a network node from a statistical viewpoint is still lacking.
Here we provide the statistical nature of $h$-index from a multivariate order statistics perspective.
In a nutshell, we can define $N$ bivariate random variables $X_{1}, X_{2},\cdots, X_{N}$ for each node where $N$ is the number of all nodes.
For each $X_{i}=(x_{i1}, x_{i2})$, the first variable records the shortest path length to the $i$-th node and the second variable corresponds to the degree of the $i$-th node.
These bivariate random variables can be ordered as $X_{(1)} \geqslant X_{(2)} \geqslant \cdots \geqslant X_{(N)}$, where $X_{(i)} \geqslant X_{(j)}$ if and only if (1) $x_{(i)1} < x_{(j)1}$ or (2) $x_{(i)1}=x_{(j)1}$ and $x_{(i)2} \geqslant x_{(j)2}$.
The standard $h$-index is the number of ranks before the $h$-th order statistic $X_{(h)}$, where $h$ is maximal value such that $x_{(h)1} = 1$ and $x_{(h)2} \geqslant h-1$.

The statistical interpretation for $h$-index enables us to obtain a new family of indices for quantifying the importance of nodes.
One road map generalizes the standard $h$-index by specifying a different order statistic: we can either use a fixed cut-off order statistic across all nodes or determine a cut-off order statistic for each node in an adaptive manner.
Reversely, we can fix a rank index and use the corresponding order statistic to assess the node importance. The above extensions share a common feature: each node is assessed based on an individual order statistic.
To fully exploit the rich information embedded in the ordered list, an ensemble approach based on rank aggregation is further presented to integrate ranking results from multiple order statistics.

Empirical studies on real networks suggest that such a statistical interpretation is reasonable and provides a new family of effective node importance indices.
Our results also unveil that the standard $h$-index is not the best one among this family and many variants are more stable and accurate.
The empirical findings, combined with our new class of measures, open the door towards using order statistics encoded in the network topology to find influential nodes.

The main contributions of our study can be summarized as follows:
\begin{itemize}
  \item The statistical nature of the $h$-index of a network node has been revealed from the perspective of multivariate order statistics.
  \item This paper presents a general framework for important node identification based on order statistics. To the best of our knowledge, this is the first work that tackles the vital node detection issue from such an aspect.
  \item Experimental results on real networks validate the effectiveness and flexibility of our method.
\end{itemize}

The remaining parts of this paper are organized as follows:
Section 2 reviews the previous works that are related to our method. Section 3 introduces our method. Section 4 presents the experimental results on 53 real networks. Finally, we draw our conclusions in Section 5.

\section{Related work}
In this section, we review previous research efforts that are closely related to our method.
In Section \ref{extentions}, we discuss the extensions of $h$-index on different types of the network.
In Section \ref{sig}, we review the existing centrality measures based on statistical inference.

\subsection{Extensions of \texorpdfstring{$h$}--index to different types of networks}\label{extentions}
Since $h$-index has been widely used for ranking nodes on unweighted networks, some researchers extend $h$-index to other types of networks.

Wu et al. \cite{weighted_h2019wu} calculate the $h$-index of each node on weighted networks via replacing the node's degree by the node's strength (the sum of weights of adjacent edges).
Yan et al. \cite{c_index2013yan} propose the $c$-index on weighted networks, where the node score can be obtained by the product of the node's strength and the strength of the neighboring node.
Based on the idea of $h$-index, Zhao et al. \cite{h_degree2011zhao} define $h$-degree on weighted networks that $h$-degree depends on computing the number of links each with strength is no less than that number.
Zhai et al. \cite{directed_h2018zhai, directed_h2019zhai} respectively introduce the $h_{in}$-index and the $h_{out}$-index on weighted directed networks, where $h_{in}$-index ($h_{out}$-index) is the $h$-index of the products of in-edge weights (out-edge weight) of the target node and the out-degrees (in-degrees) of corresponding neighbors.
Basaras et al. \cite{multilayer2017basaras} present a family of measures for computing the $h$-index on a multi-layer network.

\subsection{Centrality measures based on statistical inference}\label{sig}
Although, finding vital nodes based on statistical inference remains in its infancy.
There are some research efforts in the literature that evaluate the importance of each node from the perspective of statistical inference.

Frank \cite{frank2002bayesian} utilizes a Bayesian approach to improve the performance of node centrality evaluation methods.
Wang and Phoa \cite{wang2016focus} present the focus centrality inspired by the idea of focus test in spatial analysis, where the null hypothesis is that each node connects to its neighboring nodes with a probability that is the same as in the random network.

Furthermore, some studies identify vital nodes from biological networks using significance testing.
Liu et al. \cite{sigep2020} introduce a new method for essential protein recognition based on multiple hypothesis testing, where two important topological features including the local clustering coefficient and the degree are used as the statistics.
To reduce the running time and integrate more important properties of the real network in the null model, Liu et al. \cite{epcs2021} propose a new significance-based method in which essential protein discovery is formulated as a community significance testing problem.

In our method, we tackle the issue of vital node identification from the perspective of order statistics, which has not been investigated in previous studies.

\section{Method}

\subsection{The statistical interpretation of \texorpdfstring{$h$}--index}

Let $G=(V, E)$ denote a network with $N$ nodes $V=\{v_1, v_2, \cdots, v_N\}$ and $M$ edges $E=\{e_1, e_2, \cdots, e_M\}$.
For each node, we can define $N$ bivariate random variables $X_{1}, X_{2}, \cdots, X_{N}$. In each $X_{i}=(x_{i1}, x_{i2})$,  $x_{i1}$ is the shortest path length to node $v_{i}$ and $x_{i2}$ is the degree of $v_{i}$.
We can sort these $N$ random variables according to the following rule: $X_{i}$ is ranked before $X_{j}$ if $x_{i1} < x_{j1}$ or $ x_{i2} \geqslant x_{j2}$ when $x_{i1} = x_{j1}$.
Let $X_{(1)} \geqslant X_{(2)} \geqslant \cdots \geqslant X_{(N)}$ denote the ordered list according to the above rule (Fig. \ref{fig1} (a)).  Then, the standard $h$-index can be expressed as:

\begin{equation}
		\label{h}
		h\text{-index} = \mathop{\mathrm{argmax}}\limits_{h} |\{ {X_{(i)}|X_{(i)} \geqslant X_{(h)}, x_{(h)1}=1, x_{(h)2} \geqslant h-1} \}|,
\end{equation}
where $\mathop{\mathrm{argmax}}\limits_{h}{f(h)}$ returns the value of $h$ for which the function $f(h)$ achieves its maximal value.
According to Equation (\ref{h}), we have the following interpretation for $h$-index based on order statistics: it is the number of $X_{(i)}$s before the order statistic $X_{(h)} = \left( x_{(h)1},x_{(h)2} \right)$, where $h$ is the maximal index that can achieve both $x_{(h)1}=1$ and $x_{(h)2}\geqslant h-1$.
We take node 1 in Fig. \ref{fig1}(a) as an example to explain Equation (\ref{h}).
As shown in Tab. 1, the first row depicts order statistics derived from node 1.
The second row is the rank number of the corresponding order statistic.
Then, $h-1$ can be calculated in the third row.
Note that the order statistics (1, 4) and (1, 3) satisfy both $x_{(h)1}=1$ and $x_{(h)2} \geqslant h-1$.
Finally, the $h$-index of node 1 is 2.

\begin{figure}[!hbt]
    \centering
    \label{fig1}
    \includegraphics[width=12cm]{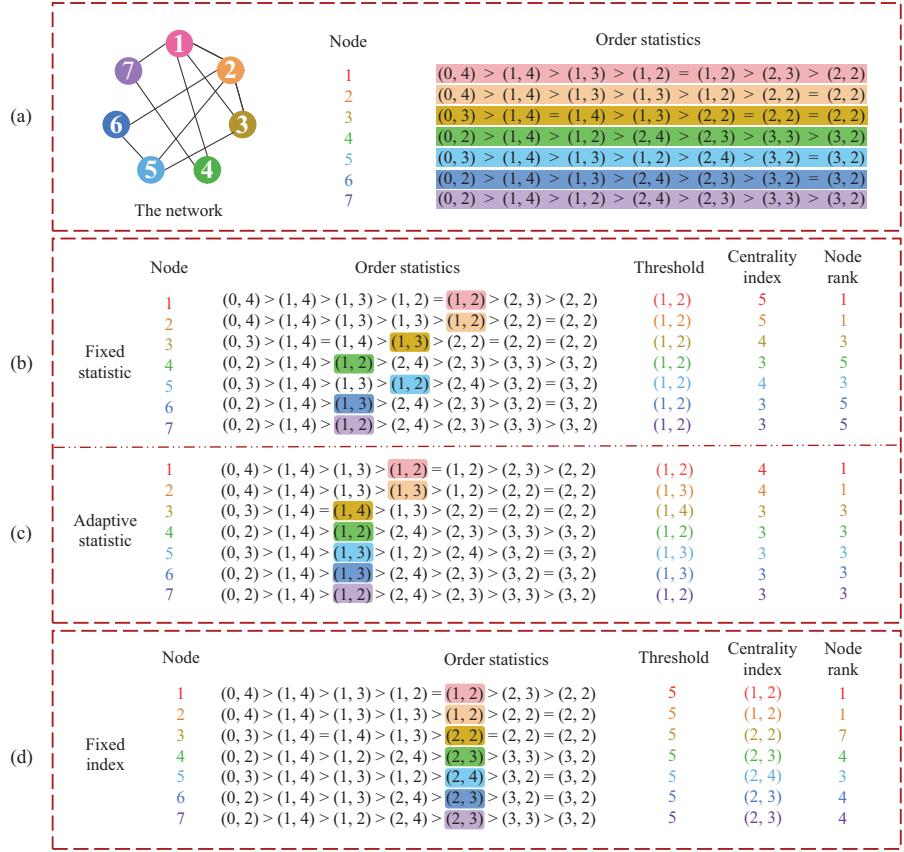}
    \caption{An example to illustrate three strategies for deriving a family of new centrality indices. (a) The input network and order statistics of each node.
    (b) An example on how to determine the centrality index based on a fixed statistic (1, 2).
    (c) An example on how to adaptively find the cut-off order statistic for each node based on the 2nd quartile.
    If the number of order statistics whose first element equals 1 is an even number, the smaller order statistic in the middle pair will be regarded as the threshold.
    (d) An example when the fixed rank index is 5 and the corresponding order statistic is used as the centrality index.
    The position of the colored rounded rectangle corresponds to rank index whose corresponding order statistic is no less than the cut-off order statistic.}
    \label{fig1}
\end{figure}

\begin{table}
    \label{h_eg}
    \includegraphics[width=12cm]{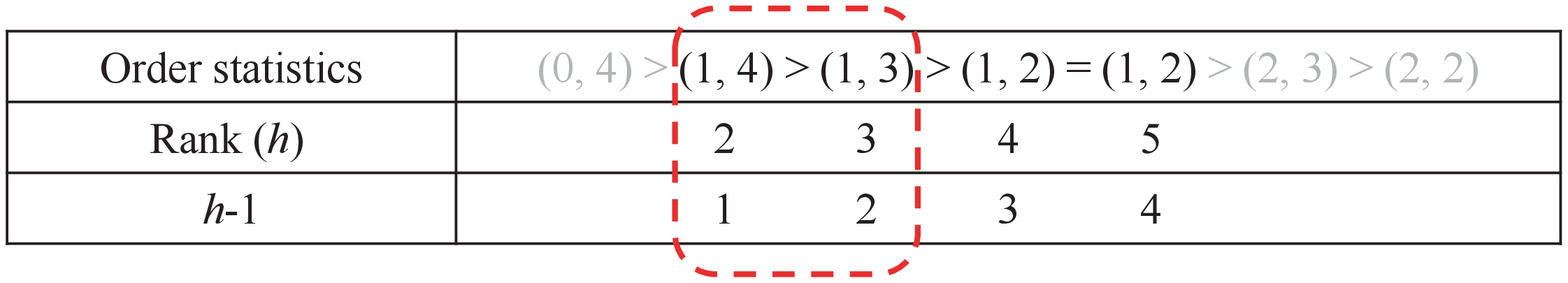}
    \caption{An example to illustrate the definition of $h$-index based on order statistics in which the $h$-index of node 1 is 2.}
\end{table}

\subsection{The generalization of \texorpdfstring{$h$}--index}

We have revealed the nature of $h$-index from the aspect of order statistics.
Obviously, we can define many alternative measures using the order statistics as well.
Hence, we can generalize the standard $h$-index to arrive at a family of new indices for evaluating the node importance.
Here we present three different strategies (Fig. \ref{fig1} (b), Fig. \ref{fig1} (c), Fig. \ref{fig1} (d)) to achieve this objective.

\subsubsection{Fixed statistic}

In $h$-index, the number of ranks before a cut-off statistic $X_{(h)}$ is used to quantify the node importance.
Obviously, we can choose a cut-off statistic based on different principles. The most simple  idea is to choose a global order statistic that is same for all nodes (Fig. \ref{fig1} (b)). Mathematically, the corresponding centrality index, denoted by FS-index, is defined as:

\begin{equation}
    \label{fs}
    FS\text{-index} = |\{ {X_{(i)}|X_{(i)}\geqslant X_{k}} \}|,
\end{equation}

\noindent where $X_{k}$ is a user-specified parameter.

According to Equation (\ref{fs}), we can obtain some interesting variants.
For instance, if we specify $X_{k} = (1,1)$, FS-index will be
\begin{equation}
\begin{aligned}
	\label{degree}
	FS\text{-index} & = |\{ {X_{(i)}|X_{(i)}\geqslant (1,1)} \}|\\
    & = |\{ X_{(i)}|x_{(i)1}=0 \} \cup \{ X_{(i)}|x_{(i)1}=1 \}|\\
    & = 1+d,
\end{aligned}
\end{equation}
where $d$ is the node's degree.
Hence, it is easy to see that the centrality index derived from the Equation (\ref{degree}) is equivalent to the Degree Centrality (DC) \cite{dc1978freeman}.
When we set $FS-index = |\{ {X_{(i)}|X_{(i)}\geqslant (1,1)} \}|-1$, it is the same as the DC.

\subsubsection{Adaptive statistic}

Different from the use of a fixed order statistic, we can alternatively determine a cut-off statistic that may be different for different nodes (Fig. \ref{fig1} (c)). Suppose we use AS-index to denote this type of centrality index, which can be expressed as:

\begin{equation}
    AS\text{-index} =  |\{ {X_{(i)}|X_{(i)} \geqslant g \left( X_{(1)}, X_{(2)}, \cdots, X_{(N)} \right) } \}|,
\end{equation}

\noindent where $g(X_{(1)}, X_{(2)}, \cdots, X_{(N)})$ is a mapping function that takes $\left( X_{(1)}, X_{(2)}, \cdots, X_{(N)} \right)$ as input and outputs a bivariate vector.
For instance, the $h$-index is a special case of such AS-index when the mapping function is defined as:

\begin{equation}
    g \left( X_{(1)}, X_{(2)}, \cdots, X_{(N)} \right)= \mathop{\arg\max}\limits_{X_{(h)}} \{ h| x_{(h)1}=1, x_{(h)2} \geqslant h-1 \}.
\end{equation}

\noindent We can use other forms of mapping functions to fulfill this task as well. Here we just present several simple possibilities that are similar to the standard $h$-index.
If we still fix the shortest path length to be 1 and compute the average degree of the 1-order neighbors, we can obtain the following mapping function:

\begin{equation}
    \label{mean}
    g \left( X_{(1)}, X_{(2)}, \cdots, X_{(N)} \right)= \left(1, \left\lceil \frac{\sum_{x_{(i)1=1}}x_{i2}}{|\{X_{(i)}|x_{(i)1}=1\}|} \right\rceil \right),
\end{equation}

\noindent where $\lceil y \rceil$ denotes the least integer greater than or equal to $y$. Similarly, we can utilize quartiles to define the mapping function as:

\begin{equation}
    g \left( X_{(1)}, X_{(2)}, \cdots, X_{(N)} \right)= (1, x_{(Q_{i})2}),
\end{equation}
where $Q_{i}$ $(i=1,2,3)$ is the $i$-th quartile of $\{1,2, ..., |\{X_{(j)}|x_{(j)1}=1\}|\}$.

\subsubsection{Fixed index}

The first two strategies use the number of ranks before an order statistic as the centrality index.
On the contrary, we can use the order statistic at a fixed rank index as the centrality index (Fig. \ref{fig1} (d)):

\begin{equation}
    FI\text{-index} = {X_{(k)}},
\end{equation}

\noindent where $1 \leqslant k \leqslant N$ is a user-specified index parameter. Different from the first two strategies, here we utilize an order statistic $X_{(k)}$ instead of the number of ranks for the purpose of node importance quantification.
First of all, according to the rule for comparing two order statistics, we are able to sort all $N$ nodes when the order statistic is used as the centrality index.
Secondly, the rationale behind this strategy is as follows.
Under the same rank index, one node will be ranked before another node if either (1) the former has a smaller shortest path length or (2) they have the same shortest path length but the former has a neighbor with a larger degree.
In both cases, the former is more likely to be an important node than the latter.

Eccentricity (ECC) \cite{ECC1995hage} can be defined as the greatest shortest path length to all nodes:
\begin{equation}
    ECC(v_i)= \max_{v_j \in V} dis(v_i, v_j),
\end{equation}
where $dis(v_i, v_j)$ is the shortest path length between $v_i$ and $v_j$.
By specifying $k = N$ in FI-index, we have
\begin{equation}
    \label{FI_N}
    FI\text{-index}=X_{(N)}=\left( x_{(N)1},x_{(N)2} \right).
\end{equation}
The ranking result based on Equation (\ref{FI_N}) is approximately consistent with that of ECC because the node with a smaller $x_{(N)1}$ will be ranked before other nodes.

\subsection{The aggregation of multiple order statistics}

Both $h$-index and its generalized indices are only dependent on one cut-off order statistic.
An appropriate cut-off order statistic is desired to improve the performance of vital node identification.
Here we present a simple solution to this issue: a cut-off parameter will be preferred if it can yield more distinct centrality index values.
Based on this idea, we can conduct parameter selection for all generalized centrality measures.

Using only one order statistic to gauge the node importance cannot fully exploit the rich information encoded in the order statistics.
To this end, we propose to integrate ranking results from multiple order statistics.
To increase the diversity, centrality indices from three different strategies are chosen.
To improve the accuracy of each component centrality index, the parameter is selected based on the method discussed above.
Note that any rank aggregation method can be utilized in our pipeline, here we choose the Stuart method \cite{stuart2003} due to its popularity in practice.

Given a normalized ordered vector $r_{(1)} \leq r_{(2)} \leq \cdots \leq r_{(l)}$ of length $l$, the Stuart method computes the joint probability $Pr \left( r'_{(1)}<r_{(1)},r'_{(2)}<r_{(2)},\cdots,r'_{(l)}<r_{(l)}\right)$, where the random vector $\left( r'_{(1)},r'_{(2)},\cdots,r'_{(l)}\right)$ is drawn independently and uniformly.
Then, the $p$-value can be calculated as:
\begin{equation}
\begin{aligned}
\label{stuart}
    Pr\left( r_{(1)},r_{(2)},\cdots,r_{(l)} \right)=& l! \int_0^{r_{(1)}} \int_{s_1}^{r_{(2)}} \cdots \int_{s_{l-1}}^{r_{(l)}} ds_l ds_{l-1} \cdots ds_1\\
    =&l! \sum_{i=1}^l \left( r_{(l-i+1)}-r_{(l-i)} \right)\times \\ &Pr \left( r_{(1)},r_{(2)},\cdots,r_{(L-i)},r_{(L-i+2)},\cdots,r_{(l)} \right).
\end{aligned}
\end{equation}
However, the calculation of the $p$-value according to Equation (\ref{stuart}) is too time-consuming, whose time complexity is $O(l!)$.
Therefore, Aerts et al. \cite{faster_stuart2006aerts} proposed a faster alternative formula with the time complexity of $O(l^2)$:
\begin{equation}
\begin{aligned}
\label{stuart_fast}
    V_k = \sum_{i=1}^{k}(-1)^{i-1} \frac{V_{k-i}}{i!}r_{(l-k+1)}^i,\\
    Pr\left( r_{(1)},r_{(2)},\cdots,r_{(l)} \right)= l!V_l,
\end{aligned}
\end{equation}
where $V_0=1$. In this paper, the $p$-value for each node is calculated based on Equation (\ref{stuart_fast}).

\begin{figure}[!ht]
    \centering
    \includegraphics[width=12cm]{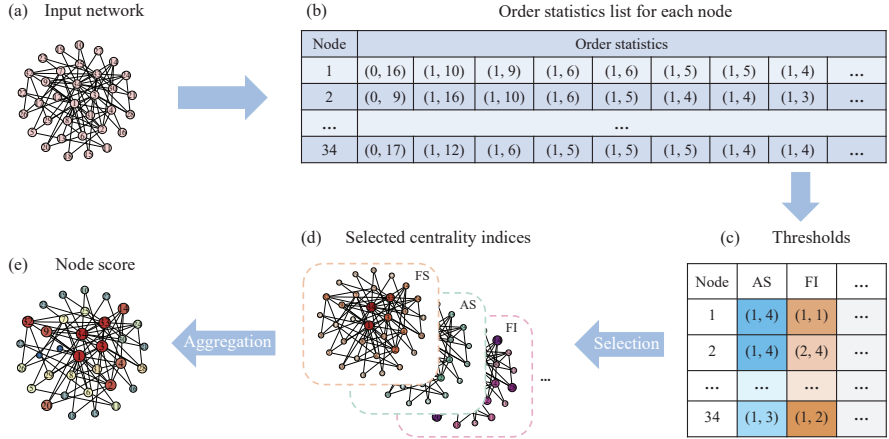}
    \caption{The main workflow for obtaining the MOS-index. (a) The input network. (b) The order statistics of each node.
    (c) Candidate thresholds based on different strategies.
    (d) Selected centrality indices based on threshold parameter selection.
    (e) Final ranking results based on the rank aggregation method.
    The size of each node is proportional to its corresponding rank.}
    \label{fig2}
\end{figure}

The method for aggregating multiple order statistics is described  in Fig. \ref{fig2}, which is denoted by MOS-index.
For a given target network in Fig. \ref{fig2} (a), we first construct the order statistics for each node shown in Fig. \ref{fig2} (b).
Then, for each strategy of FS-index, AS-index and FI-index, we specify a set of candidate cut-off order statistics or indices  (Fig. \ref{fig2} (c)), where the dark color means a bigger order statistic and the light color represents a smaller order statistic in each column.
Subsequently, the cut-off parameter selection process will be conducted to choose one threshold from each candidate set to obtain three different types of centrality indices shown in Fig. \ref{fig2} (d).
Finally, a rank aggregation method is employed to integrate selected centrality indices to obtain the consensus ranking result (Fig. \ref{fig2} (e)).
It should be noted that if the number of distinct centrality values of one selected index is smaller than others, we will omit this index from the aggregation.

\noindent The process of deriving MOS-index can be summarized as follows:

\noindent Step 1: Generate a set of order statistics for each node of the input network.\\
\noindent Step 2: Select cut-off parameters for FS-index, AS-index and FI-index based on the cut-off parameter selection strategy.\\
\noindent Step 3: Aggregate the results of FS-index, AS-index and FI-index to derive MOS-index.

\section{Results}

\subsection{Data sets}
To evaluate the performance of different indices, we conduct experiments on 53 real networks \cite{2006data} (see details in Supplementary Table S1).

\subsection{Evaluation method}

The Susceptible-Infected-Recovered (SIR) spreading model \cite{sir1992anderson} is one of the most widely used models for analyzing the propagation of epidemics and news \cite{sir2015Romualdo, sir1964daley, sir2011Iribarren}.
In the SIR spreading model, all nodes in a network can be in one of three groups, including susceptible group, infected group and recovered group.
In the initial state, there is only one seed node $v_i$ in the infected group and another $N-1$ nodes are in the susceptible group.
At each time step, each node in the infected group tries to infect their neighbors with the probability $\beta$ and then joins the group of recovered with the probability $\lambda$, where the nodes in the recovered group cannot be infected again.
This dynamic spreading process will stop when the infected group is empty.
Finally, the number of nodes in the recovered group is the spreading ability score of node $v_i$.

In this paper, the parameters in the SIR model are specified by following the common practice in previous methods \cite{lu2016h,lu2016vitalreview1,fan2021characterizing,zhou2019fast,zhang2021lfic}. More precisely, we set $\beta = \alpha \times \beta_c \approx 1.5 \times \frac{\langle k \rangle}{\langle k^2 \rangle - \langle k \rangle}$ \cite{castellano2010thresholds, newman2002threshold} and $\lambda=1$, where $\langle k \rangle$ is the average degree, $\langle k^2 \rangle$ indicates the second-order average degree, $\alpha$ is a constant that is typically set to be 1.5 in the literature such as \cite{lu2016vitalreview1,yang2021identifying}.
We use 1000 independent simulations of the SIR spreading model to obtain the final result.

The Kendall's tau correlation coefficient \cite{kendall1938new} is used for measuring the ordinal association between two ranking lists.
Let $(x_1, y_1), (x_2, y_2), \cdots, (x_N, y_N)$ be a sequence of joint ranks from two ranking lists $X$ and $Y$ respectively.
Any pair of two observations $(x_i, y_i)$ and $(x_j, y_j)$ are concordant if both elements of one observation are larger than their respective elements of the other observation.
Otherwise, they are said to be discordant.
Comparing all possible $\binom{N}{2}$ pairs, the Kendall's tau correlation coefficient can be defined as
\begin{equation}
	\tau = \frac{N_c - N_d}{\binom{N}{2}} = \frac{2(N_c - N_d)}{N(N-1)},
\end{equation}
where $N_c$ denotes the number of concordant pairs and $N_d$ represents the number of discordant pairs.

For each network, we calculate the Kendall's tau correlation coefficient between the centrality score list and the spreading ability score list obtained from the SIR spreading model \cite{sir1992anderson}.
A higher correlation coefficient means the corresponding centrality measure has better performance.

\subsection{The comparison between \texorpdfstring{$h$}--index and other variants}

We compare $h$-index with 23 centrality indices derived from three strategies.
Specifically, we respectively take (1, 2), (1, 4), (1, 6), (1, 8), (1, 10), (2, 2), (2, 4), (2, 6), (2, 8) and (2, 10) as the cut-off order statistic to evaluate the node importance according to the fixed statistic strategy;
concerning the adaptive statistic strategy, we respectively calculate $\left(1, x_{(Q_1)2}\right)$, $\left(1, x_{(Q_2)2}\right)$, $\left(1, x_{(Q_3)2}\right)$, $(1, m)$ $\left(2, x_{(Q_1)2}\right)$, $\left(2, x_{(Q_2)2}\right)$, $\left(2, x_{(Q_3)2}\right)$ and $(2, m)$ for each node, where $m$ denotes the mean of degrees calculated according to Equation (\ref{mean});
finally, we take $0.1N$, $0.2N$, $0.4N$, $0.6N$, $0.8N$ as the fixed index to find the corresponding centrality indices, respectively.

\begin{figure}[!htb]
    \centering
    \includegraphics[width=12cm]{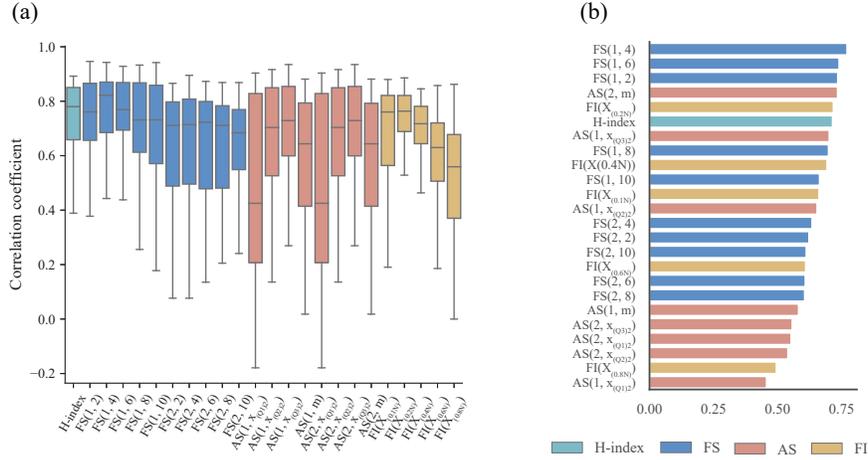}
    \caption{The comparison between $h$-index and other variants on 53 networks.
    (a) The detailed results in terms of box-plots.
    (b) The sorted average correlation coefficient.}
    \label{rq1}
\end{figure}

As shown in Fig. \ref{rq1}, the $h$-index is not the best centrality index among those variants derived from order statistics.
Some variants are more accurate and stable than $h$-index. Furthermore, centrality indices derived from different strategies have different identification performance on recognizing important nodes.
It indicates that the family of FS-index has better performance than other two strategies.
It is easy to observe that the FS-index can perform better, and it will provide better performance when the first element of the cut-off order statistic equals 1.
Therefore, we utilize (1, 2), (1, 4), (1, 6), (1, 8), (1, 10), $\left(1, x_{(Q_1)2}\right)$, $\left(1, x_{(Q_2)2}\right)$, $\left(1, x_{(Q_3)2}\right)$ and $(1, m)$ as the cut-off parameter candidates for deriving FS indices and AS indices in the following experiments.

\subsection{Cut-off order statistic selection}

\begin{figure}[!ht]
    \centering
    \includegraphics[width=12cm]{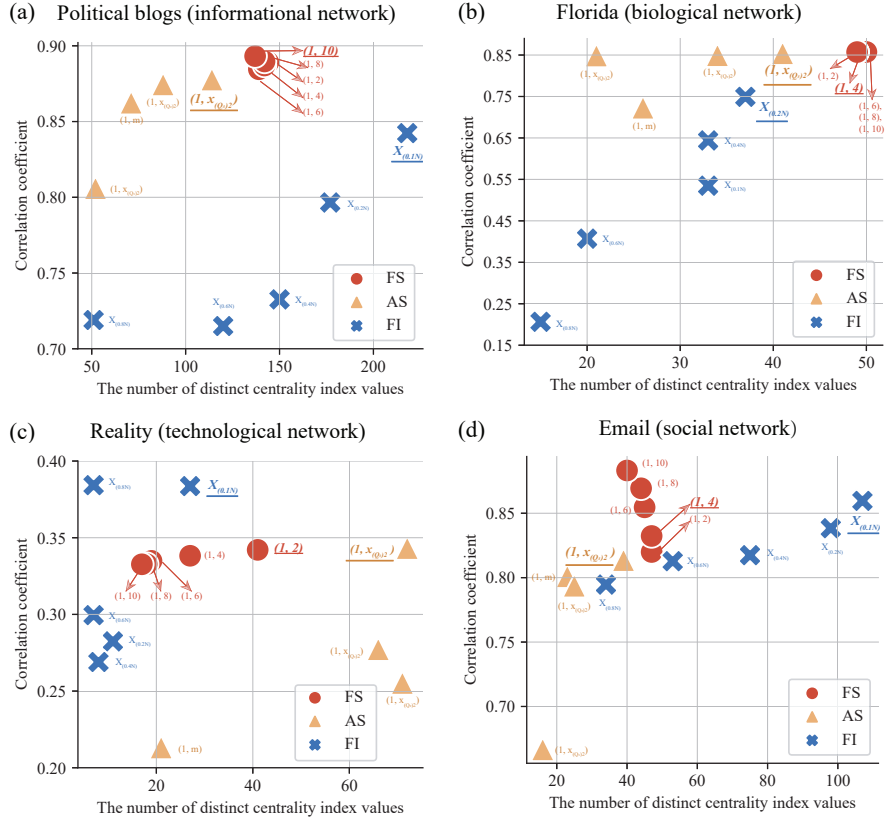}
    \caption{The cut-off parameter selection according to the number of different centrality index values on four networks.
    In each sub-figure, red, yellow and blue points represent the FS-index, AS-index and FI-index, respectively.
    The underlined order statistics are selected parameters.
    }
    \label{parameter}
\end{figure}

To choose an appropriate cut-off order statistic, we suggest using the number of distinct centrality index values as the selection criterion.
To assess its validity, we first choose four networks from four domains.
As shown in Fig. \ref{parameter}, we have marked 14 candidate order statistics for each network on the corresponding sub-figure, where the selected order statistics based on our selection criterion are underlined.
In each sub-figure, $x$-axis is the number of distinct centrality index values and $y$-axis represents the Kendall's correlation coefficient.
The red circle, yellow triangle and blue ``X" respectively describe the FS-index, AS-index and FI-index.
It is clearly visible that the order statistic with more distinct centrality index values tends to have a larger correlation coefficient.
The detailed experimental results on all 53 networks are provided in Supplementary Section 3.3.

\begin{figure}[!hbt]
    \centering
    \includegraphics[width=11.3cm]{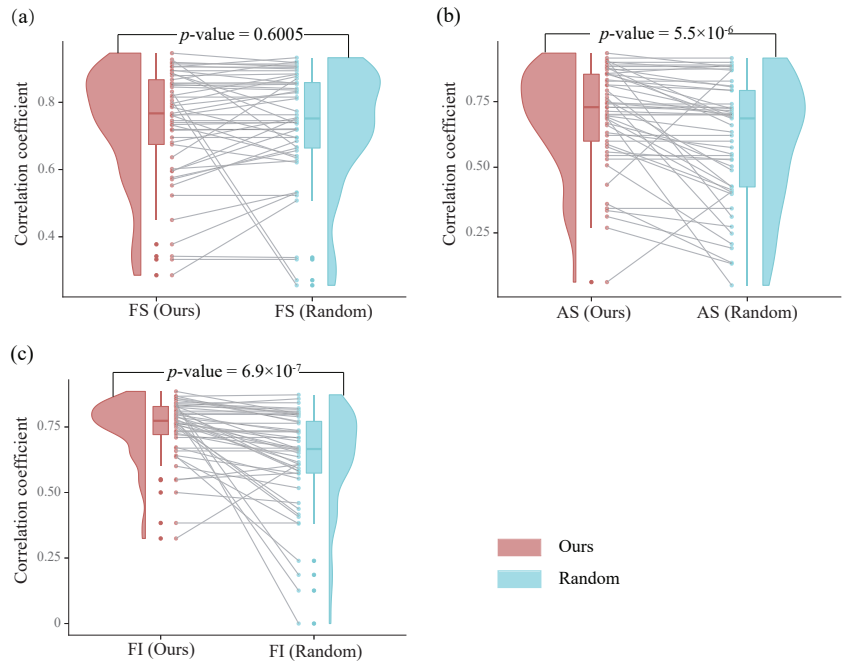}
    \caption{The comparison between our selection criterion and random selection of order statistics on 53 networks. In each sub-figure, a pair of linked up points are the experimental results on the same network.
    The light coral distribution indicates the results based on order statistics selected by our criterion while the turquoise distribution illustrates the results based on order statistics chosen by random selection.
    (a), (b), (c) The performance comparison of two parameter selection criteria based on the FS-index, AS-index and FI-index, respectively.
    }
    \label{rq4}
\end{figure}

To further demonstrate the effectiveness of our cut-off parameter selection criterion, we randomly choose a cut-off parameter from the candidate set in Fig. \ref{rq1} for each strategy.
The comparison results on 53 networks are presented in Fig. \ref{rq4}. In each of the  sub-figure, there are two distributions in which each of them is composed of 53 correlation coefficients.
From Fig. \ref{rq4} (a), Fig. \ref{rq4} (b) and Fig. \ref{rq4} (c), we can see that our criterion is superior to random selection in most cases.
Moreover, we employ the Wilcoxon signed-rank test \cite{siegel1956wilcox_test} to evaluate whether our cut-off parameter selection strategy is significantly better than random selection.
The significance  testing results in terms of $p$-values are shown in Fig. \ref{rq4}, which shows that the performance of our cut-off parameter selection criterion is significantly superior to random selection for AS-index and FI-index.

\subsection{MOS-index vs. indices based on an individual order statistic}

\begin{figure}[!bht]
    \centering
    \includegraphics[width=12cm]{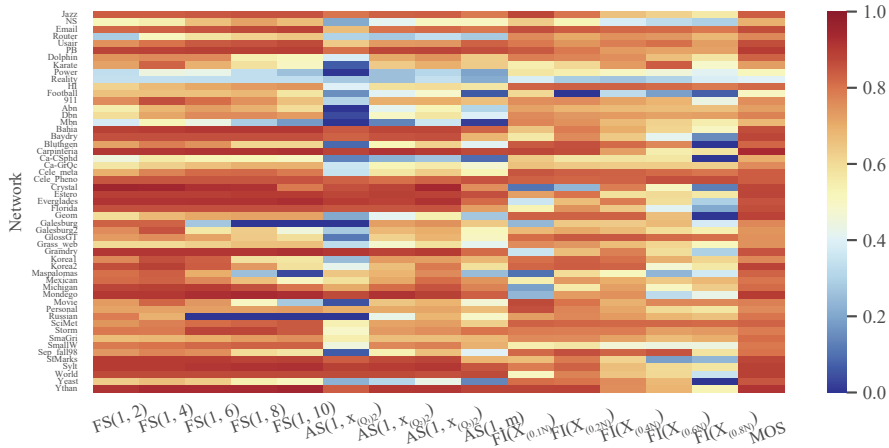}
    \caption{The comparison between 14 centrality indices and MOS-index on 53 networks.
    The dark red squares represent higher correlation coefficients while the dark blue squares indicate smaller correlation coefficients.
    The $x$-axis shows centrality indices and the $y$-axis lists all names of 53 networks.
    }
    \label{rq3}
\end{figure}

After choosing an appropriate order statistic from each strategy, the MOS-index is derived by integrating ranking lists based on these order statistics.
We compare MOS-index with indices based on 14 individual order statistics in Fig. \ref{rq3}, where MOS-index is derived by integrating three order statistics that are selected from three strategies based on our selection criterion.
In Fig. \ref{rq3}, each column represents the performance of one centrality index.
It is easy to see that the MOS-index can beat any one of 14 candidate centrality indices.

\begin{figure}[!bht]
    \centering
    \includegraphics[width=8cm]{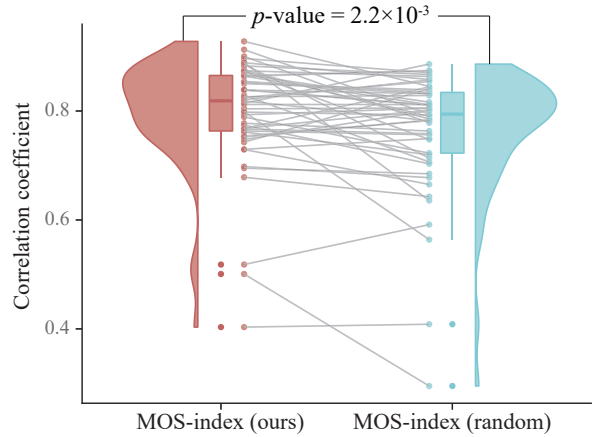}
    \caption{The results of MOS-index based on different order statistic selection criteria.}
    \label{all}
\end{figure}

Since better performance can be obtained by MOS-index, a question naturally arises as to that whether our order statistic selection criterion plays an important role in the aggregation process.
To investigate this issue, we compare the MOS-index based on our order statistic selection criterion with that based on a random selection (the same candidate set in Fig. \ref{rq1}).
According to the significance testing result in Fig. \ref{all}, MOS-index (ours) is statistically significantly better than MOS-index (random).
On one hand, this indicates that the cut-off parameter selection method is effective. On the other hand,  it demonstrates that our pipeline for calculating the aggregated MOS-index is effective as well.

\subsection{The comparison with the state-of-the-art centrality measures}
\begin{figure}[!bht]
    \centering
    \includegraphics[width=12.5cm]{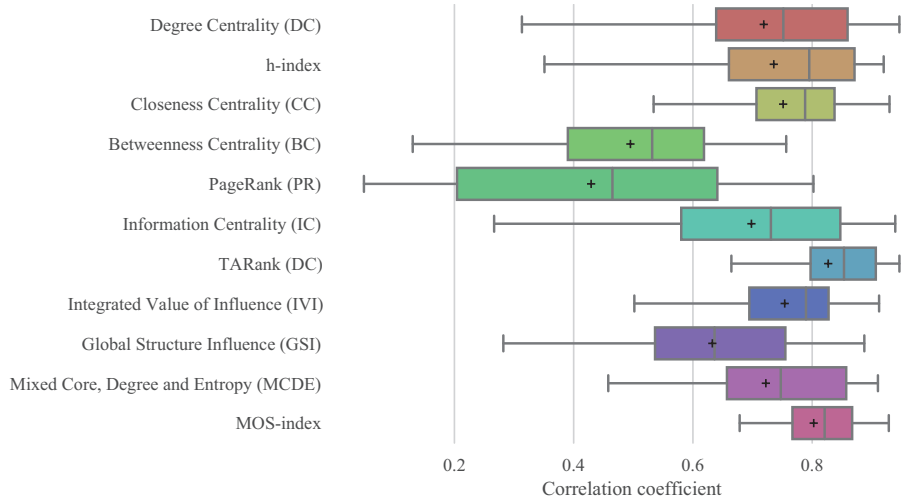}
    \caption{The comparison between MOS-index and competing centrality measures on 53 networks.
    The correlation coefficients increase from left to right in each line, the cross ``+" represents the mean value of correlation coefficients.}
    \label{rq5}
\end{figure}

We use the MOS-index as a representative of centrality indices based on order statistics and compare it with popular centrality methods and four recently proposed centrality measures: Degree Centrality (DC) \cite{dc1978freeman}, $h$-index \cite{hindex2005hirsch}, Closeness Centrality (CC) \cite{sabidussi1966CC}, Betweenness Centrality (BC) \cite{bavelas1948BC}, PageRank (PR)\cite{Brin1998pr}, Information Centrality (IC) \cite{stephenson1989IC}, TARank (DC) \cite{tarank2021}, Integrated Value of Influence (IVI) \cite{ivi2020}, Global Structure Influence (GSI) \cite{gsi2022} and Mixed Core, Degree and Entropy (MCDE) \cite{MCDE2020}.
Since TARank (DC) is not sensitive to the parameter $k$, here we set $k$ to be 2 in TARank (DC) to improve its running efficiency.
Fig. \ref{rq5} shows the comparison between MOS-index and existing centrality
measures in terms of the Kendall's tau correlation coefficient on 53 networks.

In each sub-figure of Fig. \ref{rq5}, the cross ``+" represents the mean value of correlation coefficients on 53 networks.
It is obvious that MOS-index achieves the best performance in terms of the mean value among these centrality indices except for TARank (DC).
In fact, TARank (DC) can be regarded as the aggregation of all order statistics $X_{(i)}=\left(x_{(i)1}, x_{(i)2}\right)$ that satisfy $x_{(i)1} \leqslant 1$ when $k=2$.
While the MOS-index integrates only three order statistics, which can partly explain the reason why TARank (DC) performs better than MOS-index.

\section{Conclusions}
In this article, we reveal the nature of $h$-index of a network node from an order statistic perspective.
This new angle makes it possible to generalize $h$-index to obtain a new family of centrality indices that have never been investigated.
Hence, we open the door towards assessing the importance of a network node based on the order statistics information embedded in the network.

There are still many interesting questions that remain unaddressed.
First of all, we define $N$ bivariate variables for each node, where the second component in each bivariate variable corresponds to the degree.
Since the degree can be used as a centrality index, one natural idea is to replace the degree with other existing centrality indices in the second component.
The generalization along this direction may yield many interesting variants as well.
Secondly, each bivariate variable can be generalized to be a multivariate one in which the number of components is larger than 2.
For instance, we may include an additional component variable such that it records the number of all paths of a certain length between two nodes.
Undoubtedly, the introduction of such additional information can help distinguish true vital nodes from other nodes.
Finally, we only use one or several order statistics in this article to derive centrality indices.
It can be expected that more effective measures can be obtained if all order statistics are utilized in an elegant and reasonable way.

The new family of centrality measures based on the order statistic can be extended to directed networks and weighted networks as well.
If we define the second element in the bivariate variable as the out-degree, then the corresponding order statistic can be used on directed networks.
For weighted networks, we can replace the degree by the sum of edge-weights for each corresponding node.
For other types of complex networks such as multi-layer networks and hyper-graphs, we can include more component variables to fulfill the task.

One interesting direction is to investigate the feasibility and potential of using our methodology to quantify the scientific output of a scholar.
For instance, except the number of citations for each paper, we can construct a multivariate variable for each paper with the following additional components: the number of co-authors, the reputation of published journal, the importance of the author in this paper.
Based on the constructed multivariate order statistics, we are able to better evaluate a scholar's scientific caliber than the $h$-index based solely on the number of citations.

\section*{CRediT authorship contribution statement}
\textbf{Yan Liu}: Conceptualization, Writing.
\textbf{Mudi Jiang}: Data curation, Formal analysis.
\textbf{Lianyu Hu}: Software, Visualization.
\textbf{Zengyou He}: Conceptualization, Writing, Supervision.

\section*{Acknowledgements}
\noindent This work has been supported by the Natural Science Foundation of China under Grant No. 61972066.

\section*{References}

\bibliography{mybibfile}

\begin{thebibliography}{10}
\expandafter\ifx\csname url\endcsname\relax
  \def\url#1{\texttt{#1}}\fi
\expandafter\ifx\csname urlprefix\endcsname\relax\def\urlprefix{URL }\fi
\expandafter\ifx\csname href\endcsname\relax
  \def\href#1#2{#2} \def\path#1{#1}\fi

\bibitem{albert2002statistical}
R.~Albert, A.-L. Barab{\'a}si, Statistical mechanics of complex networks,
  Reviews of Modern Physics 74 (2002) 47--97.

\bibitem{newman2003structure}
M.~E. Newman, The structure and function of complex networks, {SIAM} Review
  45~(2) (2003) 167--256.

\bibitem{LORDAN2014112traffic}
O.~Lordan, J.~M. Sallan, P.~Simo, Study of the topology and robustness of
  airline route networks from the complex network approach: a survey and
  research agenda, Journal of Transport Geography 37 (2014) 112--120.

\bibitem{LORDAN201618traffic}
O.~Lordan, J.~M. Sallan, N.~Escorihuela, D.~Gonzalez-Prieto, Robustness of
  airline route networks, Physica A: Statistical Mechanics and Its Applications
  445 (2016) 18--26.

\bibitem{morone2015social}
F.~Morone, H.~A. Makse, Influence maximization in complex networks through
  optimal percolation, Nature 524~(7563) (2015) 65--68.

\bibitem{eubank2004social}
S.~Eubank, H.~Guclu, V.~A. Kumar, M.~V. Marathe, A.~Srinivasan, Z.~Toroczkai,
  N.~Wang, Modelling disease outbreaks in realistic urban social networks,
  Nature 429~(6988) (2004) 180--184.

\bibitem{Yangeaan3184power}
Y.~Yang, T.~Nishikawa, A.~E. Motter, Small vulnerable sets determine large
  network cascades in power grids, Science 358~(6365) (2017) 860.

\bibitem{5751206power}
T.~T. Kim, H.~V. Poor, Strategic protection against data injection attacks on
  power grids, IEEE Transactions on Smart Grid 2~(2) (2011) 326--333.

\bibitem{gurney1997neural}
K.~Gurney, An introduction to neural networks, CRC press, 1997.

\bibitem{Kennedy1488Neural}
D.~Kennedy, A.~I. Selverston, M.~P. Remler, Analysis of restricted neural
  networks, Science 164~(3887) (1969) 1488--1496.

\bibitem{JACKSON2002265ecnomic}
M.~O. Jackson, A.~Watts, The evolution of social and economic networks, Journal
  of Economic Theory 106~(2) (2002) 265--295.

\bibitem{1990economic}
M.~Callon, Techno-economic networks and irreversibility, The Sociological
  Review 38~(1\_suppl) (1990) 132--161.

\bibitem{cp2021tudisco}
F.~Tudisco, D.~J. Higham, Node and edge nonlinear eigenvector centrality for
  hypergraphs, Communications Physics 4 (2021) 201.

\bibitem{cp2021fan}
T.~Fan, L.~L{\"u}, D.~Shi, T.~Zhou, Characterizing cycle structure in complex
  networks, Communications Physics 4 (2021) 272.

\bibitem{cp2022evans}
T.~S. Evans, B.~Chen, Linking the network centrality measures closeness and
  degree, Communications Physics (2022) 172.

\bibitem{lobby2009korn}
A.~Korn, A.~Schubert, A.~Telcs, Lobby index in networks, Physica A: Statistical
  Mechanics and Its Applications 388~(11) (2009) 2221--2226.

\bibitem{hindex2005hirsch}
J.~E. Hirsch, An index to quantify an individual's scientific research output,
  Proceedings of the National Academy of Sciences of the United States of
  America 102~(46) (2005) 16569--16572.

\bibitem{lu2016h}
L.~L{\"u}, T.~Zhou, Q.-M. Zhang, H.~E. Stanley, The h-index of a network node
  and its relation to degree and coreness, Nature Communications 7 (2016)
  10168.

\bibitem{weighted_h2019wu}
X.~Wu, W.~Wei, L.~Tang, J.~L{\"u}, et~al., Coreness and h-index for weighted
  networks, IEEE Transactions on Circuits and Systems I: Regular Papers 66~(8)
  (2019) 3113--3122.

\bibitem{directed_h2018zhai}
L.~Zhai, X.~Yan, G.~Zhang, Bi-directional h-index: A new measure of node
  centrality in weighted and directed networks, Journal of Informetrics 12~(1)
  (2018) 299--314.

\bibitem{directed_h2019zhai}
L.~Zhai, X.~Yan, G.~Zhang, The bi-directional h-index and b-core decomposition
  in directed networks, Physica A: Statistical Mechanics and Its Applications
  531 (2019) 121715.

\bibitem{multilayer2017basaras}
P.~Basaras, G.~Iosifidis, D.~Katsaros, L.~Tassiulas, Identifying influential
  spreaders in complex multilayer networks: A centrality perspective, IEEE
  Transactions on Network Science and Engineering 6~(1) (2019) 31--45.

\bibitem{lu2016vitalreview1}
L.~L{\"u}, D.~Chen, X.~Ren, Q.~Zhang, Y.~Zhang, T.~Zhou, Vital nodes
  identification in complex networks, Physics Reports 650 (2016) 1--63.

\bibitem{das2018review2}
K.~Das, S.~Samanta, M.~Pal, Study on centrality measures in social networks: a
  survey, Social Network Analysis and Mining 8 (2018) 13.

\bibitem{review32019Liu}
J.~Liu, X.~Liu, Z.~Hong, X.~Zeng, Q.~Zou, Y.~Lin, A.~Rodr\'{i}guez-Pat\'{o}n,
  {Computational methods for identifying the critical nodes in biological
  networks}, Briefings in Bioinformatics 21~(2) (2020) 486–497.

\bibitem{review42019Li}
M.~Li, W.~Li, R.~Zheng, X.~Li, M.~Zeng, {Network-based methods for predicting
  essential genes or proteins: a survey}, Briefings in Bioinformatics 21~(2)
  (2019) 566--583.

\bibitem{review52019bian}
R.~Bian, Y.~S. Koh, G.~Dobbie, A.~Divoli, Identifying top-$k$ nodes in social
  networks: A survey, ACM Computing Surveys 52~(1) (2019) 1--33.

\bibitem{core_structure2021wang}
R.~W. Wang, S.~X. Wei, Y.~Y. Fred, Extracting a core structure from
  heterogeneous information network using h-subnet and meta-path strength,
  Journal of Informetrics 15~(3) (2021) 101173.

\bibitem{core_structure2018zhang}
R.~J. Zhang, H.~E. Stanley, F.~Y. Ye, Extracting h-backbone as a core structure
  in weighted networks, Scientific Reports 8~(1) (2018) 1--7.

\bibitem{core_structure2014zhao}
S.~X. Zhao, P.~L. Zhang, J.~Li, A.~M. Tan, F.~Y. Ye, Abstracting the core
  subnet of weighted networks based on link strengths, Journal of the
  Association for Information Science and Technology 65~(5) (2014) 984--994.

\bibitem{core_structure2009schubert}
A.~Schubert, A.~Korn, A.~Telcs, Hirsch-type indices for characterizing
  networks, Scientometrics 78~(2) (2009) 375--382.

\bibitem{c_index2013yan}
X.~Yan, L.~Zhai, W.~Fan, C-index: A weighted network node centrality measure
  for collaboration competence, Journal of Informetrics 7~(1) (2013) 223--239.

\bibitem{h_degree2011zhao}
S.~X. Zhao, R.~Rousseau, Y.~Y. Fred, h-degree as a basic measure in weighted
  networks, Journal of Informetrics 5~(4) (2011) 668--677.

\bibitem{frank2002bayesian}
O.~Frank, Using centrality modeling in network surveys, Social Networks 24~(4)
  (2002) 385--394.

\bibitem{wang2016focus}
T.-C. Wang, F.~K.~H. Phoa, Focus statistics for testing network centrality on
  uncorrelated random graphs, Network Science 4~(4) (2016) 460--473.

\bibitem{sigep2020}
Y.~Liu, H.~Liang, Q.~Zou, Z.~He, Significance-based essential protein
  discovery, IEEE Transactions on Computational Biology and Bioinformatics
  19~(1) (2022) 633 -- 642.

\bibitem{epcs2021}
Y.~Liu, W.~Chen, Z.~He, Essential protein recognition via community
  significance, IEEE Transactions on Computational Biology and Bioinformatics
  18~(6) (2021) 2788 -- 2794.

\bibitem{dc1978freeman}
L.~C. Freeman, Centrality in social networks conceptual clarification, Social
  Networks 1~(3) (1978) 215--239.

\bibitem{ECC1995hage}
P.~Hage, F.~Harary, Eccentricity and centrality in networks, Social networks
  17~(1) (1995) 57--63.

\bibitem{stuart2003}
J.~M. Stuart, E.~Segal, D.~Koller, S.~K. Kim, A gene-coexpression network for
  global discovery of conserved genetic modules, Science 302~(5643) (2003)
  249--255.

\bibitem{faster_stuart2006aerts}
S.~Aerts, D.~Lambrechts, S.~Maity, P.~Van~Loo, B.~Coessens, F.~De~Smet, L.-C.
  Tranchevent, B.~De~Moor, P.~Marynen, B.~Hassan, et~al., Gene prioritization
  through genomic data fusion, Nature Biotechnology 24~(5) (2006) 537--544.

\bibitem{2006data}
V.~Batagelj, A.~Mrvar, Pajek datasets,
  \url{http://vlado.fmf.uni-lj.si/pub/networks/data/} (2006).

\bibitem{sir1992anderson}
R.~M. Anderson, R.~M. May, Infectious diseases of humans: dynamics and control,
  Oxford university press, New {Y}ork, 1992.

\bibitem{sir2015Romualdo}
R.~Pastor-Satorras, C.~Castellano, P.~Van~Mieghem, A.~Vespignani, Epidemic
  processes in complex networks, Reviews of Modern Physics 87 (2015) 925--979.

\bibitem{sir1964daley}
D.~J. Daley, D.~G. Kendall, Epidemics and rumours, Nature 204 (1964) 1118.

\bibitem{sir2011Iribarren}
J.~L. Iribarren, E.~Moro, Branching dynamics of viral information spreading,
  Physical Review E 84 (2011) 046116.

\bibitem{fan2021characterizing}
T.~Fan, L.~L{\"u}, D.~Shi, T.~Zhou, Characterizing cycle structure in complex
  networks, Communications Physics 4~(1) (2021) 1--9.

\bibitem{zhou2019fast}
F.~Zhou, L.~L{\"u}, M.~S. Mariani, Fast influencers in complex networks,
  Communications in Nonlinear Science and Numerical Simulation 74 (2019)
  69--83.

\bibitem{zhang2021lfic}
H.~Zhang, S.~Zhong, Y.~Deng, K.~H. Cheong, Lfic: Identifying influential nodes
  in complex networks by local fuzzy information centrality, IEEE Transactions
  on Fuzzy Systems (2021) Early Access.

\bibitem{castellano2010thresholds}
C.~Castellano, R.~Pastor-Satorras, Thresholds for epidemic spreading in
  networks, Physical Review Letters 105~(21) (2010) 218701.

\bibitem{newman2002threshold}
M.~E. Newman, Spread of epidemic disease on networks, Physical Review E 66
  (2002) 016128.

\bibitem{yang2021identifying}
X.-H. Yang, Z.~Xiong, F.~Ma, X.~Chen, Z.~Ruan, P.~Jiang, X.~Xu, Identifying
  influential spreaders in complex networks based on network embedding and node
  local centrality, Physica A: Statistical Mechanics and its Applications 573
  (2021) 125971.

\bibitem{kendall1938new}
M.~G. Kendall, A new measure of rank correlation, Biometrika 30~(1/2) (1938)
  81--93.

\bibitem{siegel1956wilcox_test}
S.~Siegel, Nonparametric statistics for the behavioral sciences, McGraw-hill,
  New {Y}ork, 1956.

\bibitem{sabidussi1966CC}
G.~Sabidussi, The centrality index of a graph, Psychometrika 31~(4) (1966)
  581--603.

\bibitem{bavelas1948BC}
A.~Bavelas, A mathematical model for group structures, Human Organization 7~(3)
  (1948) 16--30.

\bibitem{Brin1998pr}
S.~Brin, L.~Page, The anatomy of a large-scale hypertextual web search engine,
  Computer Networks \& Isdn Systems 30 (1998) 107--117.

\bibitem{stephenson1989IC}
K.~Stephenson, M.~Zelen, Rethinking centrality: methods and examples, Social
  Networks 11~(1) (1989) 1--37.

\bibitem{tarank2021}
Y.~Liu, X.~Wei, W.~Chen, L.~Hu, Z.~He, A graph-traversal approach to identify
  influential nodes in a network, Patterns 2~(9) (2021) 100321.

\bibitem{ivi2020}
A.~Salavaty, M.~Ramialison, P.~D. Currie, Integrated value of influence: an
  integrative method for the identification of the most influential nodes
  within networks, Patterns 1~(5) (2020) 100052.

\bibitem{gsi2022}
R.~D. Shetty, S.~Bhattacharjee, A.~Dutta, A.~Namtirtha, {GSI}: An influential
  node detection approach in heterogeneous network using covid-19 as use case,
  IEEE Transactions on Computational Social Systems (2022) Early Access.

\bibitem{MCDE2020}
M.~Wang, W.~Li, Y.~Guo, X.~Peng, Y.~Li, Identifying influential spreaders in
  complex networks based on improved k-shell method, Physica A: Statistical
  Mechanics and its Applications 554 (2020) 124229.

\end{thebibliography}

\end{document}